\pdfoutput=1
\def\ver{0.01} \def\verdate{06/X/2011} 
\def\ver{0.02} \def\verdate{12/I/2012} 
\def\ver{0.03} \def\verdate{16/I/2012} 
\def\ver{0.04} \def\verdate{17/I/2012} 
\def\ver{0.05} \def\verdate{19/I/2012} 
\def\ver{0.06} \def\verdate{19/I/2012} 
\def\ver{1.00} \def\verdate{20/I/2012} 
\def\ver{1.01} \def\verdate{05/III/2012} 
\def\ver{1.02} \def\verdate{06/III/2012} 
\def\ver{2.00} \def\verdate{07/III/2012} 

\let\accentvec\vec 
\documentclass[smallextended]{svjour3}

\let\vec\accentvec
\usepackage{amsfonts,amsmath}
\usepackage{dsfont}

\usepackage{amsfonts,amsmath}
\usepackage{amssymb}

\newcommand{\ket}[1]{\ensuremath{|#1\rangle}}

\newcommand{\1}{{\rm 1\hspace{-0.9mm}l}}
\newcommand{\Id}{\1}

\newcommand{\halmos}{\newline\vspace{3mm}\hfill $\Box$}

\providecommand{\proof}{\noindent {\it Proof. \ }}

\newcommand{\tr}{\ensuremath{\mathrm{tr}}}

\newcommand{\eg}{\emph{e.g.}}
\newcommand{\ie}{\emph{i.e.}}

\usepackage{tikz}
\usetikzlibrary{topaths,calc}
\def\hyperpic{
\begin{tikzpicture}
\node (v1) at (0,0) {};
\node (v2) at (2,0) {};
\node (v3) at (4,0) {};
\node (v4) at (1,1.732) {};
\node (v5) at (3,1.732) {};
\node (v6) at (2,3.464) {};
\node (m12) at (1,0) {};
\node (m24) at (1.5,0.866) {};
\node (m41) at (0.5,0.866) {};
\node (m23) at (3,0) {}; 
\node (m35) at (3.5,0.866) {}; 
\node (m52) at (2.5,0.866) {}; 
\node (m45) at (2,1.732) {}; 
\node (m56) at (2.5,2.598) {}; 
\node (m64) at (1.5,2.598) {};
\begin{scope}[fill opacity=0.7]
\filldraw[fill=yellow!70] ($(v1)+(-0.25,-0.25)$) 
    to[out=-60,in=180] ($(m12)$) 
    to[out=0,in=240] ($(v2) + (0.25,-0.25)$) 
    to[out=60,in=-60] ($(m24)$) 
    to[out=120,in=0] ($(v4) + (0,0.3535)$)
    to[out=180,in=60] ($(m41)$) 
    to[out=240,in=120] ($(v1)+(-0.25,-0.25)$);
\filldraw[fill=blue!50, thick, dotted] ($(v2)+(-0.25,-0.25)$) 
    to[out=-60,in=180] ($(m23)$) 
    to[out=0,in=240] ($(v3) + (0.25,-0.25)$) 
    to[out=60,in=-60] ($(m35)$) 
    to[out=120,in=0] ($(v5) + (0,0.3535)$)
    to[out=180,in=60] ($(m52)$) 
    to[out=240,in=120] ($(v2)+(-0.25,-0.25)$);
\filldraw[fill=red!50, dashed] ($(v4)+(-0.25,-0.25)$) 
    to[out=-60,in=180] ($(m45)$) 
    to[out=0,in=240] ($(v5) + (0.25,-0.25)$) 
    to[out=60,in=-60] ($(m56)$) 
    to[out=120,in=0] ($(v6) + (0,0.3535)$)
    to[out=180,in=60] ($(m64)$) 
    to[out=240,in=120] ($(v4)+(-0.25,-0.25)$);
\end{scope}
\foreach \v in {1,2,...,6} {\fill (v\v) circle (0.06) node [below] {$v_{\v}$};}
\node at (1,0.5) {$E_1$};
\node at (3,0.5) {$E_2$};
\node at (2,2.232) {$E_3$};
\end{tikzpicture}
}

\begin{document}

\title{Local controllability of quantum systems}
\date{\verdate \ (ver. \ver)}

\author{Zbigniew Pucha{\l}a}
\institute{Z. Pucha{\l}a \at
Institute of Theoretical and Applied Informatics, Polish Academy
of Sciences, Ba{\l}tycka 5, 44-100 Gliwice, Poland\\\email{z.puchala@iitis.pl}}

\maketitle

\begin{abstract}
We give a criterion that is sufficient for controllability of multipartite
quantum systems. We generalize the graph infection criterion to the quantum
systems that cannot be described with the use of a graph theory. We introduce the
notation of hypergraphs and reformulate the infection property in this setting.
The introduced criterion has a topological nature and therefore it is not
connected to any particular experimental realization of quantum information
processing.
\end{abstract}

\section{Introduction}

The controllability of a given quantum system is a fundamental issue of the
quantum information science. It concerns whether it is possible to drive a
quantum system into a previously fixed state.  There have been proposed
different notations of controllability of quantum systems,  such as  state
controllability, equivalent state controllability or operator controllability
\cite{albertini2001notions,albertini2003notions,d2008introduction}.  In the case
of finite dimensional quantum systems the criteria for  controllability can be
expressed in terms of Lie-algebraic concepts
\cite{albertini2002lie,d2008introduction,elliott2009bilinear}. These concepts
provide a mathematical tool, in the case of closed quantum systems, \ie{}
systems without external influences.  However, the Lie-algebraic criteria may be
difficult to check, especially when the dimension of the system in question is
large. For this reason there has been proposed methods for verifing 
controllability based on a graph theory
\cite{turinici2001quantum,turinici2003wavefunction,burgarth2007full,burgarth2009local}
and in many cases this setting provides an easier way for controllability inspection.

It is an important question whether the system is controllable when the control 
is performed only on a subsystem. This kind of approach is called
a~\emph{local-controllability} and can be considered only in the case when the
subsystems of  a given system interact. For examples may serve coupled spin
chains or spin networks 
\cite{burgarth2007full,d2008introduction,burgarth2009local}.

In \cite{burgarth2007full} there has been derived a graph infection criterion 
that ensures the controllability by relaxation and in \cite{burgarth2009local}
it has been shown, that it also can be used in the case of algebraic control. In
this paper we generalize the graph infection criterion to the quantum systems
that cannot be described with the use of a graph theory, like spin $S= 1/2$
extended $XY$ model \cite{titvinidze2003phase,lou2004thermal} or $p$-spin
interaction model \cite{goldschmidt1990solvable,baxter1973exact}. In order to
provide new criteria for controllability, we introduce the notation of
hypergraphs and reformulate the infection property in this setting. The
introduced hypergraphs infection criterion has a topological nature and
therefore it is not connected to any particular experimental realization of
quantum information processing.

This paper is organized as follows.
In Section~\ref{sec:q-mech-control-sys} we provide a general description of a
quantum mechanical control system. We introduce notation of controllability and
provide a necessary and sufficient criteria of controllability of quantum
systems.
In Section~\ref{sec:local-control} we introduce a notion of local
controllability, provide a definition of hypergraph, define hypergraph infection
property and finally give a new criterion for controllability.
In Section~\ref{sec:final} we provide the summary of the presented work and give
some concluding remarks.

\section{Quantum Mechanical Control Systems}\label{sec:q-mech-control-sys}

The dynamics of closed quantum systems can be described by a Schr\"odinger
equation
\begin{equation} \label{eqn:SchContEqn-gen}
\frac{d}{dt} \ket{\psi(t)} = 
- i H(t) \ket{\psi(t)},
\end{equation}
where $\ket{\psi(t)}$ is an element of the complex sphere $\mathrm{S}^{N-1}$ 
representing a pure state, $H(t)$ is a matrix function, which is Hermitian for
every $t$ and called a~Hamiltonian of the system. In this paper we assume, that
the Hamiltonian is in the form  $ H(t) = H_{\delta} + \sum_{i=1}^{M} h_{i}(t)
H_i$, thus a quantum system can be described as
\begin{equation} \label{eqn:SchContEqn}
\frac{d}{dt} \ket{\psi(t)} 
= -i \left(H_{\delta} + \sum_{i=1}^{M} h_{i}(t) H_i \right)\ket{\psi(t)},
\end{equation}
where matrices $H_{\delta}, H_1, H_2, \dots , H_M$ are Hermitian. Term
$H_{\delta}$ is called a drift term since it drives an evolution in no control
is applied. For given controls, the equation (\ref{eqn:SchContEqn}) is
(time-variant) linear, and thus has unique solution. In this case the system is
bilinear \cite{elliott2009bilinear}, with specialization that drift and control
matrices are skew-Hermitian.

The solution of (\ref{eqn:SchContEqn}) can be given as 
\begin{equation}
\ket{\psi(t)} = U(t) \ket{\psi(0)},
\end{equation}
where $\ket{\psi(0)}$ is an initial condition and $U(t)$ is the solution of 
an~operator equation
\begin{equation} \label{eqn:SchContOperEqn}
\frac{d}{dt} U(t) =  -i \left(H_{\delta} + \sum_{i=1}^{M} h_{i}(t) H_i \right)U(t),
\end{equation}
with an initial condition $U(0) = \1$ ($N \times N$ identity matrix). The
solution $U(t)$ is an element of Lie group of unitary matrices $\mathrm{U}(N)$,
if we assume that $\tr H_{\delta} = \tr H_i = 0$, then the solution is in a
group of special unitary matrices $\mathrm{SU}(N)$.


There are various notion of controllability for the system given in 
(\ref{eqn:SchContEqn}),
\cite{albertini2001notions,albertini2003notions,d2008introduction}. Here we will
consider two of them.
\begin{definition}[Operator Controllable System]
We call system (\ref{eqn:SchContEqn}) \emph{Operator Controllable} if it is 
possible to drive an operator $X$ in (\ref{eqn:SchContOperEqn}) to any value in 
$\mathrm{U}(N)$ (or $\mathrm{SU}(N)$).
\end{definition}

\begin{definition}[State Controllable System]
We call system (\ref{eqn:SchContEqn}) \emph{State Controllable} if it is 
possible to drive the state $\ket{\psi}$ from the complex sphere
$\mathrm{S}^{N-1}$ to  any other state on the sphere.
\end{definition}

One can also define \emph{Equivalent State Controllable System} where it is 
possible to drive any initial state to any element on the complex sphere  modulo
a phase factor, but since from a physics point of view states that differ only
by a phase factor are indistinguishable, thus the  equivalent state
controllability is equivalent to state controllability.


The main theorem concerning controllability conditions on bilinear quantum
systems follows from more general fact concerning controllability on Lie groups
and was proved in the 70's of the last century~\cite{jurdjevic1972control}. If
we specify the theorem to the bilinear quantum systems it can be stated as
follows.

\begin{theorem} \label{th:control-conditions}
Let us denote by $\mathcal{L}$ the Lie algebra generated by the matrices $i
H_{\delta}, iH_1, \dots, i H_M$, \ie{} $\mathcal{L} = \{ i H_{\delta}, i H_1,
\dots, i H_M \}_{\mathfrak{L}}$. We have the following 
\begin{itemize}
\item The system is operator controllable if and only if the Lie algebra 
$\mathcal{L}$  is an algebra $u(N)$ or $su(N)$.

\item The system is state controllable if and only if $\mathcal{L}$ is $u(N)$,
$su(N)$ or in the case of even $N$ the algebra $\mathcal{L}$ is isomorphic to 
$sp\left(\frac{N}2\right)^2$.
\end{itemize}

\end{theorem}

\section{Local controllability}\label{sec:local-control}

Let $V = C \cup \bar{C}$ be a given composite system, with Hamiltonian in the 
following form
\begin{equation}
 H = H_{\delta} + \sum_{k=1}^{M} h_k(t) H_k,
\end{equation}
where $H_{\delta} $ is a drift -- in most situations coupling Hamiltonian on whole 
system, and $H_k$ are local Hamiltonians acting on subsystem $C$, thus are in
the form $H_k = H^C_k \otimes \1_{\bar{C}}$.  The action of Hamiltonians $H_k$
are governed by time depended parameters  $h_k(t)$. By the
Theorem~\ref{th:control-conditions}, $V$ is operator controllable if and only if
$i H_{\delta}$ and $i H_k$ are generators of the Lie algebra of skew-Hermitian
operators on the composite system $V$,  \ie{}  $\{i H_{\delta}, i H_1 , i H_2, \dots, i
H_M\}_{\mathfrak{L}} = u(V)$. In this paper we assume, that the control
Hamiltonians generates the full unitary algebra on a subsystem $C$, \ie{} $\{i
H_1^C, i H_2^C, \dots, i H_M^C \}_{\mathfrak{L}} = u(C)$.


In article \cite{burgarth2009local} was given a sufficient criterion that
guarantees that a many-body quantum system with drift described by a network can
be controlled by properly manipulating the (local) Hamiltonian of one of its
subsystems. The criterion is based on a topological properties of the graph
defined by the coupling terms in a drift Hamiltonian $H_{\delta}$. The
applications of above result in the case of Heisenberg spin chains are presented
in a paper  \cite{heule2010local}.

\subsection{System specification}
Assume, that we have a composite system $\mathcal{X} = \mathcal{X}_1 \cup
\mathcal{X}_2 \cup \dots \cup \mathcal{X}_n$. The associated Hilbert space we
denote by $\mathcal{H}^\mathcal{X} =\mathcal{H}^{\mathcal{X}_1} \otimes \dots
\otimes \mathcal{H}^{\mathcal{X}_n}$. Let $i$ be an identifier of a system
$\mathcal{X}_i$ and by $X$ we denote a set of subsystems identifiers $X=\{1,2,
\dots, n\}$. The dimension of the Hilbert space $\mathcal{H}^\mathcal{X}$ is
equal to  $N = dim(\mathcal{H}^\mathcal{X}) = \prod_{i=1}^M
dim(\mathcal{H}^{\mathcal{X}_i})$.

We say, that a Hermitian operator $H$ acting on $\mathcal{H}^\mathcal{X}$ acts
only on a subsystem $i$ 
if $H$ is in the form
\begin{equation}
H = \1 \otimes \dots \otimes  \1 \otimes H_{i} \otimes  \1 \otimes  \dots
\otimes  \1,
\end{equation}
where, the operator $H_i$ acts on a $\mathcal{H}^{\mathcal{X}_i}$. More
generally we say, that a Hermitian operator $H$ acting on
$\mathcal{H}^\mathcal{X}$ acts only on a subsystem $P \subset X$ 
if 
\begin{equation}
[H, H_i ] = 0,
\end{equation}
for all operators $H_i$ acting on subsystems $i \notin P$.


As usual we denote by $u(N)$ the Lie algebra of skew-Hermitian $N \times N$
matrices. If a composite system is specified, we will use a notation
$u(\mathcal{X}_i)$ for Lie algebra of $N \times N$ skew-Hermitian matrices
acting on  $\mathcal{H}^{\mathcal{X}_i}$. If $i$ is an identifier of a
subsystem $\mathcal{X}_i$, we will write $u(i)$ to denote this algebra.
Similarly if $P \subset X$, by $u(P)$ we denote a Lie algebra of $N \times N$
skew-Hermitian acting on a~subsystem~$P$.

\subsection{Hypergraphs -- definitions and properties}

In this section we give a definition of a hypergraph and provide an infection 
property on a hypergraph. The infection property for hypergraphs has been
adopted from papers \cite{burgarth2007full,burgarth2009local}, where it was
defined for graphs and used to provide controllability conditions for quantum
networks.

\begin{definition}[Hypergraph]
Let $X = \{x_1, x_2 , \dots , x_n\}$ be a finite set of nodes, and let 
$\mathcal{E} = (E_1, E_2, \dots , E_k)$ be a family of subsets of $X$, we will
call them edges. If we have, that $E_i \neq \emptyset$ for $i=1,2,\dots,k$ and
$\cup_{i=1}^k E_i = X$, then we call $(X,\mathcal{E})$ a \emph{hypergraph}
(see~\cite{berge1976graphs}). In this paper we assume, that $X = \{1, 2 , \dots
, n\}$.
\end{definition}

The infection process on a hypergraph can be stated as follows: if some set of
nodes $A$ is \emph{infected}, then the infection spreads onto a \emph{healthy} 
neighbours $B$ if there exists edge $E$ joining some elements of infected group
$A_1 \subset A$ with $B$, $E=A_1 \cup B$. Moreover $E$ is the only edge that
joins elements from $A_1$ with healthy nodes. 

\begin{definition}[Infection spread]
Assume that $A \subset X$ is infected, we say that infection can spread onto $B
\subset X$ from $A$, where $A \cap B = \emptyset$ if
\begin{equation}
\exists_{E \in \mathcal{E}} 
E \cap A \neq \emptyset \text{ and } E \setminus A = B,
\end{equation}
moreover if $x\in(E \cap A)$ and $x \in F$ for some $F \in
\mathcal{E}$, then $F=E$ or $F \subset A$.
\end{definition}
If there exists an initial set of nodes $C$ that can infect whole hypergraph,
we call such set infecting.
\begin{definition}[Hypergraph infection property] \label{def:hypgr-inf}
For a hypergraph $(X,\mathcal{E})$ we call a subset $C \subset X$ infecting if
there exist a sequence  $C = P_1 \varsubsetneq P_2 \varsubsetneq \dots
\varsubsetneq P_m = X$ such that an infection can spread from $P_i$ onto $P_{i+1}
\setminus P_{i}$ for $i=1,2,\dots,m-1$.
\end{definition}

\begin{example} \label{ex:hyper-ex}
To illustrate the above definitions we provide an example. Let us define a
hypergraph $(X,\mathcal{E})$ with nodes $X=\{1,2,\dots,6\}$ and edges
$\mathcal{E} = \{E_1,E_2,E_3\} =  \{\{1,2,4\},$ $\{2,3,5\},$ $\{4,5,6\}\}$. The graphical
representation of the hypergraph is presented in fig.~\ref{fig:hyper-graph}. 
\begin{figure}[h]
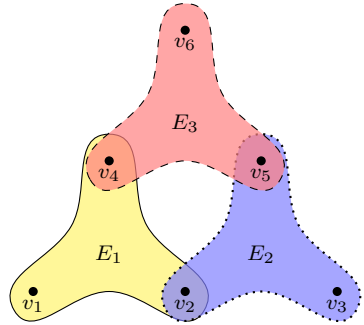
 \label{fig:hyper-graph}
\centering 
\hyperpic 
\caption{
Graphical representation of hypergraph with nodes $X=\{1,2,\dots,6\}$ and edges
$\mathcal{E} = \{E_1,E_2,E_3\}= \{\{1,2,4\},$ $\{2,3,5\},$ $\{4,5,6\}\}$
}
\end{figure}

If one assumes, that a set $A = \{1,2,4\}$ is infected, then it is easy to see,
that the infection can spread \eg{} onto $\{3,5\}$ by an edge $E_2=\{2,3,5\}$.
Similarly if a set $\{1,2,3,4,5\}$ is infected, then infection can spread onto
$\{6\}$ by an edge $E_3=\{4,5,6\}$. The above gives us, that the hypergraph
$(X,\mathcal{E})$ has an infection property, with an infecting set $C=\{1,2,4\}$
and a sequence $C=P_1 =  \{1,2,4\} \varsubsetneq P_2 = \{1,2,3,4,5\}
\varsubsetneq P_3= \{1,2,3,4,5,6\} = X$.

\end{example}

\subsection{Propagation property}

To provide a controllability conditions we must assume that the drift
Hamiltonian meets some criteria. To do so we introduce a notion of
\emph{propagating property}, which relates the underlying hypergraph with
appropriate Lie algebras.

\begin{definition}[Propagating property]
Assume, that we have a composite system $\mathcal{X} = \mathcal{X}_1 \cup
\mathcal{X}_2 \cup \dots \cup \mathcal{X}_n$. Let $(X,\mathcal{E})$ be a
hypergraph. The nodes of the hypergraph are the subsystems identifiers 
$X=(1,2, \dots, n)$. The edges are related to the Hermitian operator $H$, which
acts on a $\mathcal{H}^\mathcal{X}$. The relation is given by
\begin{equation}
H = \sum_{E \in \mathcal{E}} H_{E},
\end{equation}
where $H_{E}$ are Hermitian operators acting on subsystem $E$.
We say, that
$H$ has a \emph{propagating property} if for all $E \in \mathcal{E}$,
and for all $\emptyset \neq E' \subset E$, we have
\begin{equation}
\{[i H_{E} , u(E')], u(E')\}_{\mathfrak{L}} = u(E).
\end{equation}
\end{definition}

\begin{example} \label{ex:hyper-ex2}
Assume that we have a hypergraph $(X,\mathcal{E})$ described in
Example~\ref{ex:hyper-ex}. We also assume, that a composite system is composed
with six qubits and for an edge $E = \{e_1,e_2,e_3\}$ of the hypergraph the
Hermitian operator $H_E$ is given by
\begin{equation}
i H_{E} = 
S_2^{e_1} S_2^{e_2} S_2^{e_3} +
S_3^{e_1} S_3^{e_2} S_3^{e_3} +
S_1^{e_1} S_2^{e_2} S_3^{e_3} +
S_2^{e_1} S_1^{e_2} S_0^{e_3} +
S_3^{e_1} S_2^{e_2} S_0^{e_3},
\end{equation}
where $S_j^{m}$ is $j^{\text{th}}$ Pauli matrix on
subsystem $m$, in this case
\begin{equation}
S_j^{(m)} = \underbrace{\Id \otimes \dots \otimes \Id}_{m-1} \otimes \sigma_j
\otimes \underbrace{\Id \otimes \dots \otimes \Id}_{6-m},
\end{equation}
with notation $\sigma_0 = \Id$.  The Hamiltonian $H = \sum_{E \in \mathcal{E}}
H_{E}$ has a propagation property. It is quite cumbersome task to check this
analytically, but it can be done using computer algebra systems with symbolic
computation and a~procedure to generate a basis of a dynamical Lie algebra, see
\eg{} \cite[Chapter 3.2.1]{d2008introduction}. Since the propagation property
must be checked only on a small subsystem, this computation is fast and
efficient.

\end{example}

\subsection{Main theorem}

Now we can state the main theorem, which express a new controllability criterion
in notion of hypergraph infection and propagating property. The proof of the
theorem follows the similar line of argument that the proof of the theorem
in~\cite{burgarth2009local}.
\begin{theorem} \label{th:main-th}
Let us assume, that the drift Hamiltonian $H_{\delta}$ has a propagating
property, we also assume, that a subset $C \subset X$ infects the hypergraph
$(X,\mathcal{E})$. Then, the system is controllable if we perform, the control 
only on a subsystem~$C$.
\end{theorem}
\proof First we will show by induction, that $u(P_j) \subset \{i
H_{\delta},u(C)\}_{\mathfrak{L}}$, where subsets $P_j$ are defined in the
hypergraph infection property (Def.~\ref{def:hypgr-inf}). The first induction
step is obvious, since  $u(P_1) = u(C) \subset \{i
H_{\delta},u(C)\}_{\mathfrak{L}}$. Next, we assume, that for some $j < m$, we
have
\begin{equation}
u(P_j) \subset \{i H_{\delta},u(C)\}_{\mathfrak{L}}.
\end{equation}
Since, we have, that infection spreads from $P_{j}$ onto $P_{j+1} \setminus
P_{j}$, there exist an edge $E \in \mathcal{E}$, such that  $E \cap P_j \neq
\emptyset$ and $E \setminus P_{j} = P_{j+1} \setminus P_{j}$.

We define $E^{(j)} = E \cap P_j$ and write 
\begin{equation}
\begin{split}
[i H_{\delta} , u(E^{(j)})] &= 
\sum_{F \in \mathcal{E}} [i H_{F}, u(E^{(j)})] 
\\ &= [i H_{E} , u(E^{(j)})] 
+ 
\sum_{F \in \mathcal{E}, F\neq E} [i H_{F}, u(E^{(j)})].
\end{split}
\end{equation}
Now using, the infection property of the graph $(X,\mathcal{E})$, we have, that 
the last sum above can be restricted to $F \in \mathcal{E}, F \subset P_j$. Thus
we have
\begin{equation}
[i H_{E} , u(E^{(j)})] = 
[i H_{\delta} , u(E^{(j)})]  -
\sum_{F \in \mathcal{E}, F \subset P_j} [i H_{F}, u(E^{(j)})].
\end{equation}
The last sum is an element of $u(P_j)$ and the first term of the right hand side
is an element of $\{i H_{\delta},u(C)\}_{\mathfrak{L}}$, so we obtain, that $[i
H_E, u(E^{(j)})] \in \{i H_{\delta}, u(C)\}_{\mathfrak{L}}$.

Now using propagation property of the drift Hamiltonian, we have 
\begin{equation}
\{[i H_{E} , u(E^{(j)})], u(E^{(j)})\}_{\mathfrak{L}} = u(E).
\end{equation}
Since $\{u(P_j),u(E)\}_{\mathfrak{L}} = u(P_{j+1})$, thus $u(P_{j+1}) \in
\{i H_{\delta},u(C)\}_{\mathfrak{L}}$.

At the last step of induction procedure $j = k$, we obtain 
\begin{equation}
u(P_k) = u(X) \subset \{i H_{\delta}, u(C)\}_{\mathfrak{L}},
\end{equation}
and since $u(X)$ is the maximal algebra which can be obtained, we have 
\begin{equation}
u(P_k) = u(X) = \{i H_{\delta},u(C)\}_{\mathfrak{L}}.
\end{equation}
Now by Theorem \ref{th:control-conditions}, we obtain that the system is 
operator controllable. 
\halmos

To use the above theorem to check whether the system is controllable one first
must test the propagation property for a drift Hamiltonian and ensure that the
underlying hypergraph has an infection property. The second task is purely
topological and can be done rather easily, while the first one -- the
propagation property must be checked only on a small subsystem, which reduces
the complexity.  

Using Theorem~\ref{th:main-th} we obtain that the dynamical system described in
Examples~\ref{ex:hyper-ex}, \ref{ex:hyper-ex2} is controllable by performing the
control only on a subsystem $\{1,2,4\}$. We remind here the assumption, which
was made at the beginning of Section~\ref{sec:local-control}, that local control
Hamiltonians generates the full unitary algebra on a specified subsystem.

\section{Concluding remarks}\label{sec:final}

In this paper we provided a new controllability criterion  of multipartite
quantum systems based on notion of hypergraph. We generalized the graph
infection criterion to the quantum systems that cannot be described with the use
of a graph theory. To do so we have introduced a notion of local
controllability, provided a definition of hypergraph, defined hypergraph
infection property and finally gave a new criterion for controllability. The
introduced criterion has a topological nature and therefore it is not connected
to any particular experimental realization of quantum information processing.

\section*{Acknowledgements} 

We acknowledge the financial support by the  Polish National Science Centre
under the grant number N~N514~513340.


\end{document}